\documentclass[showpacs,10pt,twocolumn,prl]{revtex4-1}
\makeatletter

\newcommand{\Rmnum}[1]{\expandafter\@slowromancap\romannumeral #1@}
\makeatother
\usepackage{amsmath}
\usepackage{amssymb}
\usepackage{graphicx}
\usepackage{graphics}
\usepackage{epsfig}
\usepackage{CJK}
\usepackage{color}
\usepackage{epstopdf}
\usepackage{multirow}
\usepackage{tabularx}
\begin{document}

\title{Defect-tuned colossal cryo-thermopower in FeSb$_2$.}
\author{Qianheng Du$^{1,2,\star}$, Lijun Wu$^1$,  Huibo Cao$^3$, Chang-Jong Kang$^4$, Christie Nelson$^5$, Gheorghe Lucian Pascut$^{4,6}$,  Tiglet Besara$^{7,8,\dag}$, Kristjan Haule$^4$, Gabriel Kotliar$^1,4$, Igor Zaliznyak$^1$, Yimei Zhu$^1$ and Cedomir Petrovic$^{1,2,\star}$}
\affiliation{$^1$ Condensed Matter Physics and Materials Science Department, Brookhaven National Laboratory, Upton 11973 New York USA\\
$^2$ Materials Science and Chemical Engineering Department, Stony Brook University, Stony Brook 11790 New York USA\\
$^3$ Neutron Scattering Division, Oak Ridge National Laboratory, Oak Ridge, Tennessee 37831, USA\\
$^4$ Department of Physics and Astronomy, Rutgers University, Piscataway, New Jersey 08854, USA\\
$^5$ National Synchrotron Light Source II, Brookhaven National Laboratory, Upton, New York 11973, USA\\
$^6$ MANSiD Research Center and Faculty of Forestry, Stefan Cel Mare University (USV), Suceava 720229, Romania\\
$^7$ Department of Chemical and Biochemical Engineering, FAMU-FSU College of Engineering, Tallahassee, Florida 32310, USA\\
$^8$ National High Magnetic Field Laboratory, Florida State University, Tallahassee, Florida 32310, USA}

\date{\today}

\begin{abstract}
FeSb$_{2}$ hosts higher known values of thermopower in nature. By changing the amount of defects in different FeSb$_2$ crystals here we show that thermopower maxima change between relatively small 14 $\mu$V/K and colossal values of about 20 mV/K. Defects also cause rather weak but important monoclinic distortion of the crystal structure $Pnnm$$\rightarrow$$Pm$. Whereas defects of both Fe and Sb atoms are observed, Sb-defficiency at Sb11 sites of $Pm$ space group is significant, revealing the source of the in-gap states that govern thermopower size. The broad distribution of the in-gap states of Fe orbital character gives rise to thermopower consistent with the electronic diffusion mechanism but only a well-defined in-gap state drives the colossal enhancement above the electronic difusion limit. {\color{red}The absence of Sb along [010] promotes stronger quasi 1D Fe $d$ orbital overlap, which induces more anisotropy in the electronic structure and triggers the quasi-1D conductivity in optics and electricity.}
\end{abstract}

\maketitle

Thermoelectric materials exploit thermoelectric effect where temperature difference is converted into electric power and vice versa \cite{Bell,Snyder}. Progress in cryo-thermoelectric materials discovery has been much slower when compared to materials used above the room temperature \cite{Heremans,He,Herbert}. This is because at low temperature additional complexity arises since electronic correlations cannot be neglected \cite{Tomczak,Palsson,KoshibaeW}. Also, in the figure of merit $ZT = S^2 \sigma T/\kappa$ where $T$ is temperature, $S$ is thermopower and $\sigma$ ($\kappa$) are electrical (thermal) conductivities, cryogenic temperature $T$ is inevitably small \cite{Snyder}. Therefore, a cryo-thermoelectric material must maximize its thermoelectric power factor ($S^{2}$$\sigma$) where thermopower provides considerable contribution.

The iron diantimonide FeSb$_2$ is a narrow-gap semiconductor that features colossal thermopower and thermoelectric power factor at cryogenic temperatures \cite{Cedomir,Cedomir2,Bentien,JieQ}. It has been proposed that colossal thermopower in FeSb$_2$ is of electronic origin \cite{Sun,Sun1}. Consequently multiband electronic correlation are important for the enhancement of the diffusion thermopower up to about $\sim$ 1 mV/K, much higher than expected from the Mott formula $S=-[\frac{(\pi k_B)^2}{3e}]T(\partial ln\sigma/\partial E)_{E_F}$ where values of the order of 10 $\mu$V/K are commonly observed \cite{Cutler,Tomczak1}. Recently it has been argued that enhancement above the diffusion limit into values of the order of $\sim$ 10 mV/K is due to phonon drag \cite{Tomczak1,Takahashi}. {\color{red}Although the importance of the in-gap states to the thermopower was recognized, the origin of the in-gap states and the role it takes in determining the thermopower remain controversial \cite{Tomczak1,Matsuura}. The possible origin from the impurities in the Fe and Sb metals was studied and the electrical transport below 40 K shows sensitivity to even ppm-level impurities \cite{Takahashi1}. While the sample-dependence electrical resistivity was explained, the effect of these impurities to the large Seebeck coefficient and the origin of the quasi-1D conductivity in optics and electrical transport were not well understood \cite{Cedomir,Homes}.}

\begin{figure}
  \centering
  \includegraphics[width=0.45\textwidth]{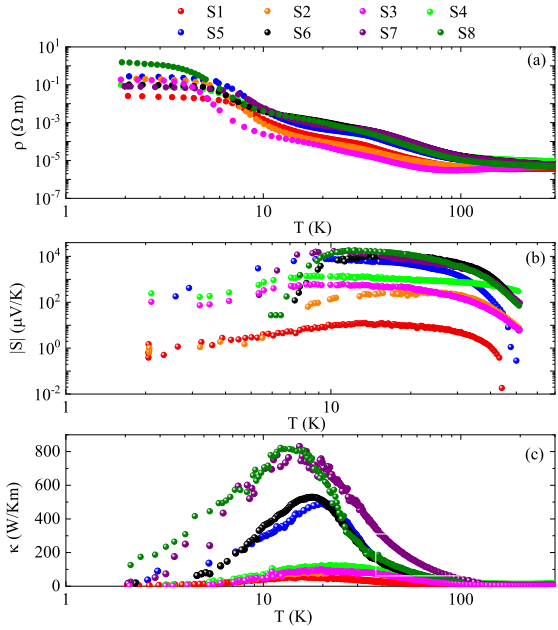}\\
  \caption{Temperature dependence of the {\color{red}(a) electrical resistivity, (b) thermopower and (c) thermal conductivity} for FeSb$_{2}$ crystals S1 - S8 with different defect content (see text).}
\end{figure}

Whereas atomic defects have been extensively used to enhance high-temperature thermoelectric performance by lowering phonon thermal conductivity in the figure of merit \cite{Biswa,KimK}, tiny amount of defects are not expected to distort crystal structure and dramatically tune thermopower values \cite{Takahashi1}. In this work we show that in correlated electron thermoelectric FeSb$_{2}$, colossal thermopower arises from Sb vacancy-induced changes in the crystal structure and associated in-gap states {\color{red}through the phonon-drag effect}\cite{Battiato}. Moreover, our results indicate that thermopower {\color{red}and the quasi-1D behavior in conductivity} are tunable by Sb defects.

Figure 1(a-b) presents electronic and thermal transport difference among eight crystals of FeSb$_2$ grown under somewhat different conditions \cite{Supplement} . Crystals S7 and S8 have about 1-2 order of magnitude higher electrical resistivity when compared to crystals S1-S3 in temperature region (10 - 20) K [Figure 1(a)]. Moreover, crystals S1-S3 have clear weak (semi)metallic resistivity in 60 K to 300 K temperature region. Low-temperature thermopower $S$ shows large variation [Fig. 1(b)]; $|$$S$$|$ maxima change by several orders of magnitude from S1 (14 $\mu$V/K) to S8 (20 mV/K). {\color{red}Figure 1(c) shows thermal conductivity $\kappa$ vs temperature for samples S1-S8. The $\kappa$ increase more than 8 times from S1 to S8 which reflects the increase scattering of phonons and the decrease of the phonon mean free path \cite{Supplement}.}

Thermally activated resistivity above 50 K [Figure 1(a)]  stems from the intrinsic energy gap consistent with the electronic structure calculations; however the resistivity shoulder in the region of $S_{max}$ around 10 K is a fingerprint of the defect-induced in-gap impurity states{\color{red} \cite{Battiato,Sun3,Takahashi2}.} Recently proposed physical mechanism of colossal thermoelectricity in FeSb$_2$ links the in-gap states to phonon-drag effect {\color{red}\cite{Battiato,Matsuura}}. {\color{red}The effects of defects in scattering the phonons and decreasing the phonon-drag effect are discussed in the supplemental material. Although defect-induced in-gap states are necessary to the large Seebeck coefficient through phonon-drag effect, excessive defects is harmful to achieve this lrage Seebeck coefficient.} The magnetoresistance (MR) changes in FeSb$_2$ should arise only in the temperature range where the in-gap band provides dominant contribution to electronic transport; a single in-gap band gives only one peak in MR \cite{Battiato}. In agreement with previous observation \cite{Sun3}, MR in high-thermopower crystals S6 and S8 shows one peak associated with a well-defined in-gap state [Figure 2(a)]. This is in contrast with MR in low-thermopower crystals S1 and S3 where broad MR is observed, consistent with a rather broad distribution of multiple in-gap states \cite{Battiato}. Another important observation is that the driver of colossal thermopower-related in-gap states is unlikely to be in Fe-rich Fe$_{1+x}$Sb$_{2-x}$ as postulated in Ref. 20. High thermopower crystal S8 shows the absence of the Curie tail in contrast to low-thermopower crystal S3 [Fig. 2(b)]. By fitting the tail of S3 using the Curie law $\chi_{ave}=\chi_0+C/T$ [Fig. 2(c)], concentration of excess Fe is about 4 atomic percents. The in-gap states emerge from different cause, as we show below.

\begin{figure}
  \centering
  \includegraphics[width=0.45\textwidth]{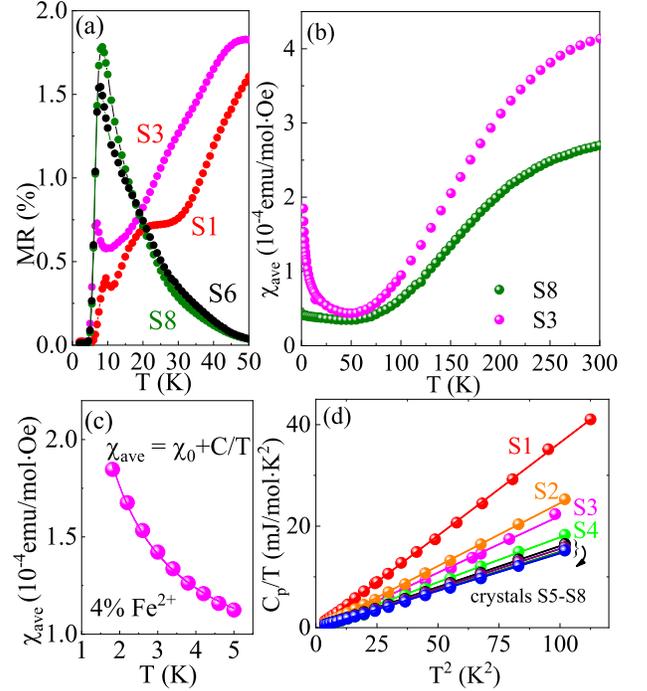}\\
  \caption{ (a) Magnetoresistance of low- (S1, S3) and high-thermopower samples (S6, S8). (b) Comparison of magnetic susceptibility of S3 and S8. (c) Low-temperature $\chi(T)$ of S3 shows clear sign of Fe impurities. (d) Low-temperature heat capacity for crystals S1-S8.}
\end{figure}

\begin{figure*}
  \centering
  \includegraphics[width=0.95\textwidth]{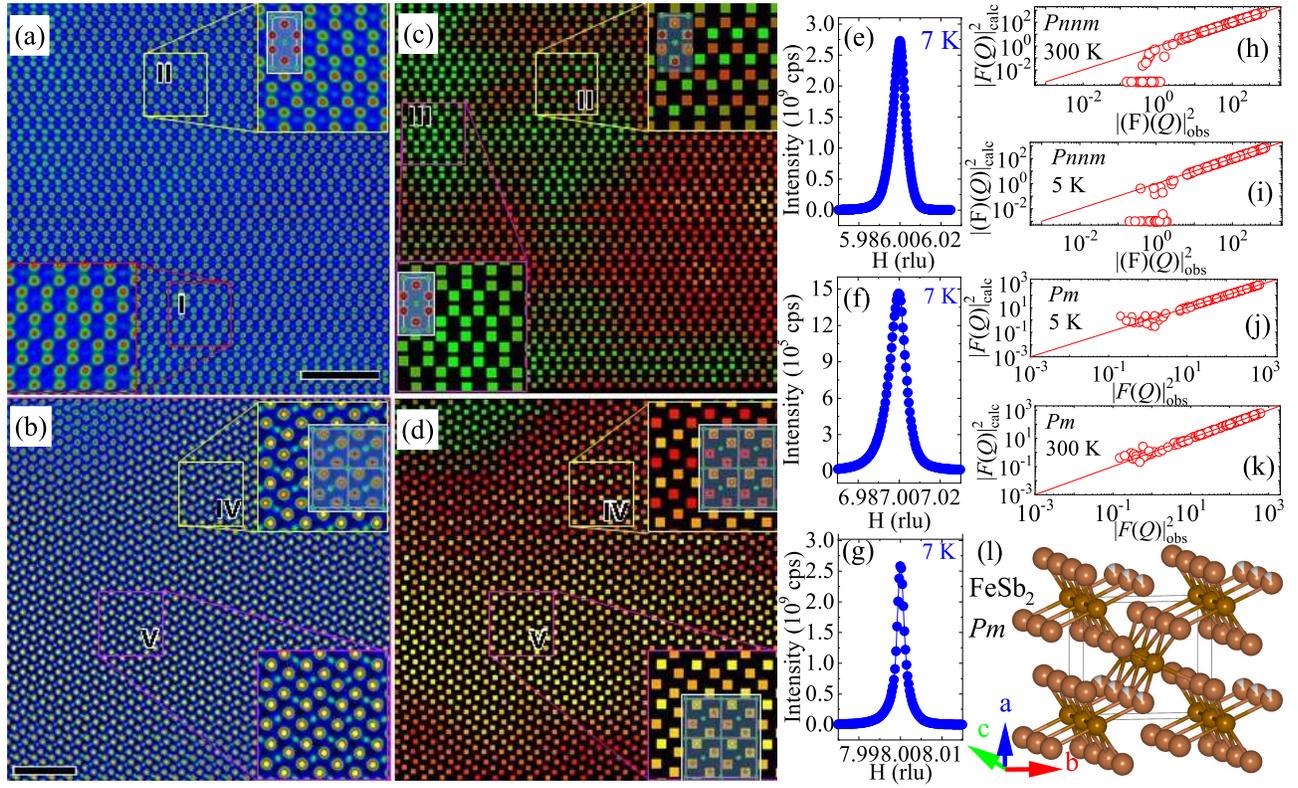}\\
  \caption{Atomic defects in FeSb$_2$ and weak crystal structure distortion. (a,b) STEM-HAADF image of low thermopower crystal S3 (a) viewed along $Pnnm$ [100] direction and high thermopower crystal S8 viewed along $Pnnm$ [001] direction (b). The insets are the magnified images from area I and II in (a) [area IV and V in (b)] with the $Pnnm$ [100]  in (a) [$Pnnm$ [001] projection in (b)] projection of the structure embedded. Scale bar 2 nm. The contrast is approximately proportional to Z$^{1.7}$ along the atomic column, thus the dots with strong and weak contrast correspond to Sb and Fe column, respectively. Red and green spheres represent Sb and Fe, respectively. (c,d) Sb peak intensity maps refined from (a) and (b), respectively. Each square represents a Sb column with intensity increasing in black-blue-green-orange-yellow-white order. The insets are magnified maps from area II-V. Note, the smooth Sb peak intensity oscillation from top to bottom in (b) and (d) could be attributed to thickness variation. (e-g) Synchrotron X-ray diffraction scans of high thermopower crystal. $Pnnm$-forbidden peak is observed at [100] wavevector (f), about 3 orders of magnitude weaker than nearby Bragg peaks. $Pnnm$ (h,i) and $Pm$ (j,k) unit cell refinements obtained in single crystal neutron diffraction experiment on low thermopower crystal, confirming structural distortion in both low- and high-thermopower material. (l) $Pm$ unit cell of FeSb$_{2 }$ induced by atomic defects. Central octahedral atom Fe (dark) is surrounded by Sb (light); see also Figure S1. The defects are preferably on Sb11 atomic sites (white arrow-marked light balls).}
\end{figure*}

\begin{figure}
  \centering
  \includegraphics[width=0.45\textwidth]{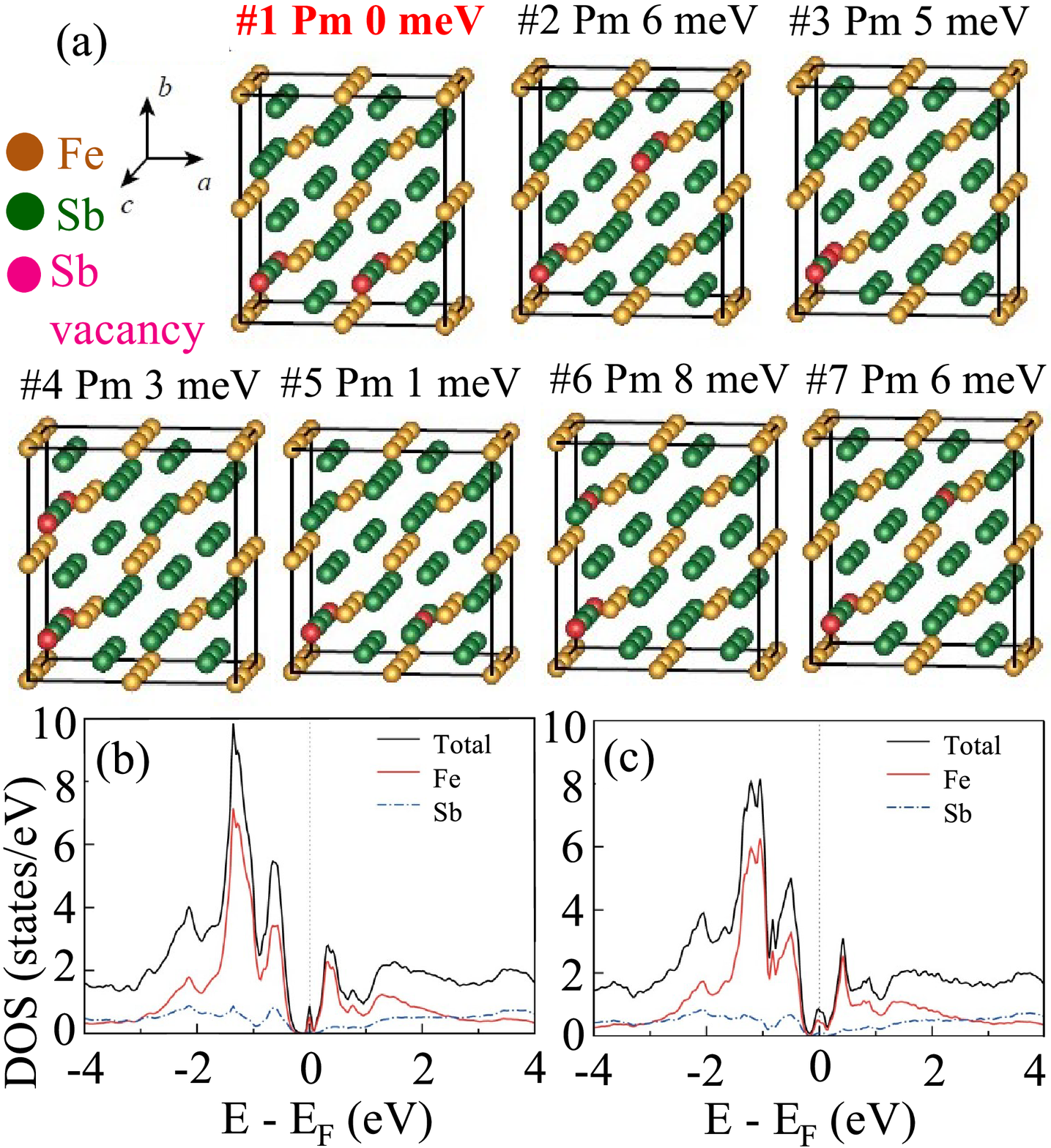}\\
  \caption{(a) Different arrangements of two Sb vacancies in the supercell and their energetics. The arrangement consistent with the $Pm$ space group has the lowest energy, in agreement with experiment. Projected density of states for (b) one and (c) two Sb vacancies. Notice that the antimony vacancy gives rise to an iron impurity band.}
\end{figure}

Figure 2(d) shows heat capacity of S1-S8 crystals. We observe a general trend of the slope change associated with Debye temperature. In general, vacancy formation energy is related to Debye temperature: $T_D=C\frac{E_V^{1/2}}{M^{1/2}\Omega^{1/3}}$, where $C$ is a constant, $E_v$ is the vacancy formation energy, M and $\Omega$ are atomic mass and volume \cite{Mukherjee,March}. Low vacancy formation energy is consistent with higher number of vacancies for low thermopower crystals (e.g. S1) when compared to high thermopower crystals (e.g. S8). Results are summarized in Table S1 \cite{Supplement}.

To shed light on the vacancy defects we performed high resolution scanning transmission electron microscopy (STEM) with a high angle annular dark field (HAADF) detector as its contrast is proportional to Z$^{1.7}$ along the atom column, where Z is the atomic number. Figures 3 (a,b) show the STEM-HAADF images taken from low thermopower S3 and high thermopower S8 samples, respectively. The strong and weak dots in the images correspond to Sb and Fe atoms, respectively. The atomic arrangement in the images is consistent with the FeSb$_2$ structure with $Pnnm$ symmetry, as shown in the insets where the atomic projections are embedded in the magnified image. Due to the Z-contrast nature of the STEM-HAADF image, the peak intensity of each dot can be used to count the atoms along the column \cite{LeBeau}. Higher peak intensity indicates more atoms along the column, while weaker peak intensity indicates less atoms, thus more vacancies along the column. It is seen that the peak intensity of Fe is quite uniform in the S8 [Fig. 3(b)], indicating the relative uniform distribution of Fe. However, the peak intensity of Fe varies in S3, e.g. there are less Fe in area I than that in area II, as shown in the magnified image in the insets of Figure 3(a). We also observe Sb intensity variation, indicating the variation of Sb occupancies. The Sb occupancies or vacancies can be better resolved by refining each Sb column peak with the second order polynomial function. Fig. 3(c,d) shows the peak intensity maps of Sb for S3 and S8, respectively. There are four Sb atoms in a FeSb$_2$ unit cell. In sample S3, the Sb peak intensity is the same in some area [inset III in Fig. 3(c)], consistent with the $Pnnm$ symmetry. In the other area, however, the Sb peak intensity changes periodically, and orders along $Pnnm$ [010] direction, as shown in the inset II in Fig. 3(c), where two Sb peaks (orange squares) are stronger than the other two Sb (green squares) within the unit cell. This indicates the ordering of Sb vacancies and reduction of the $Pnnm$ symmetry in this area. In [100] projection orange and red spheres represent high and low Sb occupancy, embedded in the inset II. The size of the phase separation clusters with Sb vacancy ordering is about a few nanometers. These clusters are similar to the nanoprecipitates observed in PbTe-AgSbTe$_2$ system \cite{Hsu,Ke}. They induce additional phonon scattering, thus reduce the thermal conductivity and phonon mean free path of S3. For crystal S8, the Sb vacancy ordering is observed in nearly all locations and also along [010] direction. In area IV [Fig. 3(d)], two Sb peaks (orange) are stronger than the other two Sb (red). While in the area V, one Sb peak (yellow square) is stronger than the other three Sb (orange squares).

Sb vacancy ordering results in a $Pnnm$-forbidden peak observed in single crystal synchrotron X-ray diffraction [Figure 3(e-g)], consistent with weak structural distortion observed in neutron diffraction measurements [Figure 3 (h-k)].  Neutron diffraction  shows presence of weak (h,0,l), h + l = odd and (0,k,l), k + l = odd type Bragg reflections, which are forbidden in $Pnnm$ crystal structure previously refined for the stoichiometric FeSb$_2$ \cite{Cedomir,Cedomir2,Hagg}. The presence of these forbidden reflections hints that the symmetry of the crystal lattice is lower. Comprehensive structural refinement [Figure 4(h-k), Tables S2-S5 \cite{Supplement}] indicate that one of the two Sb sites, which are equivalent in $Pnnm$ space group shows displacive monoclinic distortions and site deficiency. This induces very subtle change in the structural symmetry ($Pnnm$$\rightarrow$$Pm$), making two Sb sites inequivalent, i. e. Sb1 site is fully occupied while Sb11 site contains substantial number of vacancies that do not change with temperature [Fig. 3 (l)].  We find that the Sb11 site is 0.82(2) occupied while both Fe sites are 0.94(2) occupied; the chemical vacancies do not change with temperature whereas intensity of $Pnnm$-forbidden peaks at 300 K is 2/3 of that at 5 K \cite{Supplement}.

The $Pnnm$ FeSb6 octahedra are edge-sharing along the shortest lattice parameter 3.194 $\AA$. Closer inspection of $Pm$ unit cell shows that the Sb11 atomic sites labeled with arrows [Fig. S1] are also separated by the lattice parameter length 3.194 $\AA$ along the edge-sharing octahedral direction. However, weak metallic conductivity at high temperature for low thermopower crystals [Figure 1(a)] is along the orthogonal, 6.536 $\AA$ lattice parameter direction where distance of Sb11 to Fe is shorter [$b$-direction in Fig. 3(l), see also Fig. S1] \cite{Cedomir,Supplement,Hu1}.  This is consistent with TEM-observed vacancy order direction [Figure 3 (a-d)] and quasi 1D conductivity in optics \cite{Perucchi,Homes}. It should be noted that occupancy of Fe $d$ orbital ($d^n$) translates into different unit cell parameter along the $d$ orbital overlap. Higher (lower) occupancy corresponds to larger (shorter) unit cell parameter \cite{Hull,Hulliger}. Detailed comparison of sample S3 and S8 crystallography by laboratory single crystal X-ray diffraction shows longer $b$ lattice parameter in low thermopower sample S3 (Table S6 \cite{Supplement}), implying differences in Fe $d$ orbital occupation. A picture emerges where Sb11 vacancies create Fe-derived conducting in-gap states due to short Fe-Sb11 hopping distance (Fig. S1 \cite{Supplement}). Low vacancy formation energy, i.e. higher vacancy content in low-thermopower crystals such as S3 [Fig. 2(d)], promotes stronger quasi 1D Fe $d$ orbital overlap due to the absence of Sb along [010], weak metallicity and metal-insulator transition \cite{JieQ}. As we show below, first-principle calculations support this scenario.

GW+DMFT calculations show that bands associated with quasi 1D dispersion along 6.536 $\AA$ are Fe-derived: bottom of the conduction band is dominated by Fe xy orbital whereas top of the valence bands is dominated by Fe xz/yz bands \cite{Homes}. First-principle calculation results [Fig. 4(a)] confirm that structure distortion to $Pm$ space group is energetically favorable in Sb-defficient $Pnnm$ FeSb$_2$ unit cell. However, the Sb vacancies also induce Fe dangling bonds that might influence electronic structure though conducting impurity band at the Fermi level for high Sb defect concentration.

To explore how Sb vacancy and its ratio affect the electronic structure of FeSb$_2$, we studied two cases of Sb vacancy ratios: one and two Sb vacancies in a 2 $\times$ 2 $\times$ 3 super-cell of FeSb$_2$. To simulate the experimental situation, we only consider Sb vacancies at the Sb11 site. Hence, Sb vacancies for both cases occupy the Sb11 site only, and they correspond to 91.7 and 83.3 $\%$ of Sb11 occupancy, respectively, where the case of two Sb vacancies is close to the experimental Sb11 occupancy, 82 $\%$ at T = 300 K. For the case of one Sb vacancy, there is only one symmetrically non-equivalent configuration for introducing one Sb vacancy at the Sb11 site in the 2 $\times$ 2 $\times$ 3 super-cell. On the other hand, for the case of two Sb vacancies, there are 7 symmetrically non-equivalent configurations. We have considered all of the 7 configurations and found that two Sb vacancies is the configuration that is the most energetically stable [Fig. 4(a)]. From now on, we only focus on the most stable configuration for each Sb vacancy ratio.

Figure 4(b) shows the density of states (DOS) of the most stable configurations for both Sb vacancy ratios. The Sb vacancies give rise to Fe dangling bonds that lead to metallic impurity bands at the Fermi level. The impurity bands have a dominant $d$-orbital character of Fe possessing the dangling bonds. We would like to note that the DFT calculation with the modified Becke-Johnson (mBJ) exchange potential method gives a clear bulk gap of $\sim$0.25 eV in FeSb$_2$ without any vacancies. As depicted in Fig. 4(b), the case of two Sb vacancies shows broader bandwidth of the metallic impurity band than that of one Sb vacancy, indicating that as the Sb vacancy ratio increases, the metallic impurity band is more dispersive. The case of one Sb vacancy does not show the quasi-1D conductivity. Therefore, we conclude that larger Sb vacancy content induces more anisotropy in the electronic structure and triggers the quasi-1D conductivity.

Now, we turn our attention into thermoelectric power of FeSb$_2$. In the experiment, the maximum value of thermoelectric power decreases as the energy of vacancy formation is lowered, i.e. with higher number of vacancies [Fig. 1(a), Fig. 2(d)]. This is consistent with a two-band model study with an ionized impurity donor state \cite{KangCJ}; the maximum value of thermoelectric power decreases as the impurity concentration increases. Hence the increased Sb vacancies decline thermoelectric power, in agreement with experiment. On the other hand we note that the presence of some small number of Sb vacancies is necessary for creation of in-gap states important in phonon drag [Fig. 2(a,d)], Fig. 3(a,b), Fig. 4 {\color{red}\cite{Tomczak1,Takahashi,Battiato,Matsuura}.}

In summary, we have examined FeSb$_2$ crystals with different amount of Sb crystal lattice defects. The defects induce rather weak distortion of the unit cell symmetry that has profound effect on the thermoelectricity due to creation of the in-gap states with Fe $3d$ orbital character. Giant thermopower arises in the delicate balance of electronic diffusion associated with Sb defect-induced in-gap states with Fe $d$ orbital character and phonon drag of such states to colossal values. The drag is efficient only for a small number of such states, which must be sufficient to create in-gap states and increase the thermopower over high (1-2) mV/K diffusion values and at the same time not be overwhelming since superabundant defects reduce both diffusion \cite{KangCJ} and phonon-drag via additional phonon scattering (Fig. 3). This reveals a subtle interplay of phonon and electronic diffusion mechanisms and points to crystallographic origin of relevant materials physics that could be exploited in predictive thermoelectric materials design.

This work was supported by the U.S. Department of energy, Office of Science, Basic Energy Sciences as a part of the Computational Materials Science Program (QHD, CJK, LP, KH, GK and CP). This research used resources at the High Flux Isotope Reactor, a DOE Office of Science User Facility operated by the Oak Ridge National Laboratory.  This research used beamline 4-ID of the National Synchrotron Light Source II, a U.S. Department of Energy (DOE) Office of Science User Facility operated for the DOE Office of Science by Brookhaven National Laboratory under Contract No. DE-SC0012704. A portion of this work was performed at the National High Magnetic Field Laboratory, which is supported by National Science Foundation Cooperative Agreement No. DMR-1157490 and the State of Florida.

$\star$petrovic@bnl.gov and qdu@bnl.gov
$\dag$ Present address: Department of Physics, Astronomy, \& Materials Science Missouri State University, Springfield, MO 65897.

\end{document}